\documentclass[prd,twocolumn,nofootinbib]{revtex4}

\usepackage{graphicx, epsfig}
\usepackage{color}
\usepackage{mathrsfs}
\usepackage{bm}
\usepackage{amsmath,amssymb,empheq}
\usepackage{mathrsfs}
\usepackage[caption=false]{subfig}
\usepackage{hyperref}
\usepackage[normalem]{ulem}
\usepackage{mathtools}
\usepackage[x11names]{xcolor}
\usepackage{orcidlink}

\definecolor{fashionfuchsia}{rgb}{0.96, 0.0, 0.63}
\colorlet{no_so_fashion_purple}{blue!50!red}

\newcommand{\be}{\begin{equation}}
\newcommand{\ee}{\end{equation}}
\newcommand{\ba}{\begin{eqnarray}}
\newcommand{\ea}{\end{eqnarray}}

\newcommand{\nn}{\nonumber}
\newcommand{\half}{\frac{1}{2}}

\def\half{\frac{1}{2}}

\def\Tr{{\rm Tr}}
\def\rmdiag{{\rm diag}}

\def\SU(3){{\rm SU(3)}}
\def\SUthree{{\rm SU(3)}}
\def\SU(2){{\rm SU(2)}}
\def\U(1){{\rm U(1)}}
\def\U(2){{\rm U(2)}}
\def\Z2{{\rm Z}_2}

\begin{document}
\title{Domain walls and magnetic monopoles in Grand Unified Models}
\author{Harish Hemming\footnote{hhemming@asu.edu}, 
Tanmay Vachaspati\footnote{tvachasp@asu.edu}, 
Anja Wachowitz\,\orcidlink{0009-0005-0920-379X}\footnote{anja.wachowitz@asu.edu}
}
\affiliation{
Physics Department, Arizona State University, Tempe,  Arizona 85287, USA. 
}

\begin{abstract}
Motivated by Grand Unification, we study the formation of magnetic monopoles in an SU(3) non-Abelian gauge theory. We find that the number density of magnetic monopoles depends critically on a parameter, $\epsilon$, that controls the abundance and subsequent decay of biased domain walls. For sufficiently small but non-vanishing values of $\epsilon$, very few monopoles and walls survive in our simulations, potentially solving the cosmological monopole over-abundance problem.
In addition, the scenario predicts a stochastic gravitational background from biased domain walls and the possibility of magnetically charged black holes.
\end{abstract}

\maketitle

The quest for a Grand Unified Theory (GUT) of the fundamental forces suffers a setback
because all GUT models predict the existence of magnetic monopoles whereas none have
been seen. Based on simple estimates, the predicted abundance of magnetic 
monopoles far exceeds cosmological constraints~\cite{Vilenkin:2000jqa}.
A possible solution to this cosmological magnetic monopole
problem is an inflationary epoch that occurs after GUT symmetry breaking and that
vastly dilutes the abundance of magnetic monopoles in the universe. 
However, 
the formation of magnetic monopoles during GUT symmetry breaking has never been directly
verified, and it is possible that details of the symmetry breaking may themselves yield
a solution to the cosmological magnetic monopole problem. Such a scenario, known
as ``monopole sweeping'', was envisaged in Ref.~\cite{Dvali:1997sa} and certain
aspects of the scenario have been further explored in Refs.~\cite{Pogosian:1999zi,Pogosian:2000xv,Vachaspati:2001pw,Pogosian:2001fm,
Pogosian:2002ua,Antunes:2003be,Vachaspati:2003zp,Brush:2015vda,Vachaspati:2006zz,
Dvali:2022rgx,Bachmaier:2023zmq,Senjanovic:2025enc,Bachmaier:2026}. The idea is that GUT symmetry breaking may result
in the formation of magnetic monopoles and biased domain walls. The interactions of
monopoles and domain walls may result in the effective annihilation of monopoles,
greatly reducing their cosmological abundance. We wish to study this process by directly studying
the process of GUT symmetry breaking, the production of magnetic monopoles and
domain walls, their evolution, and the final density of magnetic monopoles as a
function of parameters of the GUT model.

The minimal GUT is based on an SU(5) symmetry group and, at a minimum, involves 
120 bosonic fields.
(We ignore fermionic fields for the purposes of our numerical simulations.) Such a huge
number of fields poses computational problems and it is desirable to find a smaller model
which can replicate the physics. For this reason we focus on an SU(3) gauge field theory
with a scalar in the adjoint representation of SU(3). The model is still quite large as it has 
8 scalar and 24 gauge fields (in temporal gauge) and requires another 8 variables to implement
Gauss constraints, making for a total of 40 field variables. The Lagrangian is,
\be
L = \Tr ([D_\mu, \Phi]^2) - \half \Tr (W_{\mu\nu}^2) - V(\Phi )
\label{lagrangian}
\ee
where $D_\mu = \partial_\mu - ig W_\mu$ and $W_{\mu\nu} = [D_\mu,D_\nu]/(-ig)$.
The generators of SU(3) are taken to be the Gell-Mann matrices, 
denoted $T_a$, normalized by $\Tr (T_a T_b) = \delta_{ab}/2$.
The potential function is chosen to be
\ba
V(\Phi ) &=& - m_2 \Tr (\Phi^2) + \epsilon \Tr(\Phi^3) + \lambda \Tr(\Phi^4) \nn \\
&& \hskip 0.5 cm
+  \lambda_6 (\Tr(\Phi^2))^3  + d_6 (\Tr(\Phi^3))^2
\ea
It is not necessary to include other operators up to dimension 6 
because of the SU(3) relations,
\ba
 \det (\Phi) &=& \Tr(\Phi^3) /3,  \ \ 
(\Tr (\Phi^2) )^2 = 2 \Tr (\Phi^4) , \nn \\ 
 (\Tr(\Phi^2))^3 &=& 4\Tr(\Phi^6) -4 (\Tr(\Phi^3))^2 /3. \nn
\ea
We introduced order $\Phi^6$ terms in the potential because it is known that without
them the potential has O(8) symmetry and the symmetry breaking pattern is not uniquely 
picked out. To clarify this further, $\Phi$ can get a vacuum expectation value (VEV) in the
$T_3 = \rmdiag (1,-1,0)/2$ direction, breaking SU(3) to ${\rm U(1)} \times {\rm U(1)} / \Z2$, or it can get
a VEV in the $T_8=\rmdiag(1,1,-2)/(2\sqrt{3})$ direction which would break SU(3) to
$\U(2)= \SU(2) \times {\rm U(1)} /\Z2$. The resulting vacua are degenerate in energy if we 
only include quartic 
order operators in the potential. The order 6 terms in the potential break the degeneracy 
and, with a suitable choice of parameters, yield a VEV in the $T_8$ direction, breaking 
SU(3) to U(2), which is similar to GUT symmetry breaking.

The $\Tr (\Phi^3)$ term in the potential is introduced to explicitly break the 
$\Z2$ symmetry under $\Phi \to -\Phi$. If this term is absent ($\epsilon=0$), the $\Z2$ 
symmetry breaking gives
topological domain walls, while the SU(3) symmetry breaking gives magnetic monopoles.
The interaction of domain walls and magnetic monopoles is the mechanism underlying the
sweeping of magnetic monopoles~\cite{Dvali:1997sa}. With a non-zero but small $\epsilon$, the sweeping
mechanism should still work, and the (biased) domain walls will eventually annihilate as the
vacua across them are not degenerate~\cite{Larsson:1996sp}. We will study monopole
formation and evolution as a function of the parameter $\epsilon$.

The first step is to determine parameters for which the symmetry
breaking pattern is $\SUthree \to \U(2)$. Any $\Phi$ can be diagonalized to lie in the 
$T_3-T_8$ plane and so we restrict attention to $\Phi = \phi_3 T_3 + \phi_8 T_8$, insert this
form into $V(\Phi)$ and extremize $V$ as a function of $\phi_3$ and $\phi_8$.
Further analysis yields the parameter space for which
the minima occur in the $T_8$ vacua. An example of such parameters is,
\be
m_2=\half, \ \epsilon =0, \ \lambda = \frac{3}{4}, \ \lambda_6 = 1, \ d_6 = -5.9.
\label{parameters}
\ee
With these parameters $\eta^2 /2 \equiv \Tr(\Phi^2) \approx 0.64$ in the true vacuum.
In the $\phi_3-\phi_8$ plane the 6 minima that occur are in directions
$\pm T_8$, $\pm \rmdiag(1,-2,1)/2\sqrt{3}$ and $\pm \rmdiag(-2,1,1)/2\sqrt{3}$. 
With $\epsilon=0$, the $+$ and $-$ vacua are degenerate. If we
take $\epsilon>0$, the minima with the $+$ signs become lower than those with the $-$ signs.

\begin{figure}
\includegraphics[width=0.40\textwidth,angle=0]{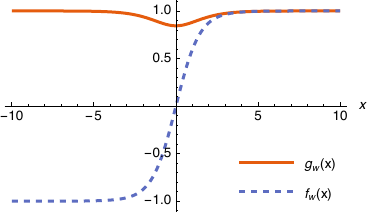}
 \caption{Domain wall profile functions $f_w(x)$ and $g_w(x)$.}
\label{dwprofiles}
\end{figure}

Let us first consider $\epsilon=0$, when the $Z_2$ symmetry is exact. Then a domain wall
interpolates between $\Phi = \eta \, \rmdiag(1,-2,1)/2\sqrt{3}$ and 
$\Phi = - \eta \, \rmdiag(-2,1,1)/2\sqrt{3}$, among other possibilities~\cite{Vachaspati:2006zz}. 
The explicit solution may be written as,
\be
\Phi_{\rm dw} (x)  = \eta \left ( g_w(x) \frac{\sqrt{3}}{2} T_3 + f_w(x) \frac{1}{2} T_8 \right )
\label{Phidw}
\ee
where $f_w(\pm \infty) = \pm 1$ , $g_w'(0)=0$, $g_w(\pm \infty)=1$. Fig.~\ref{dwprofiles} shows a 
plot of the profile functions. Note that $\Phi_{\rm dw}(x=0) \propto T_3$.

\begin{figure}
\includegraphics[width=0.40\textwidth,angle=0]{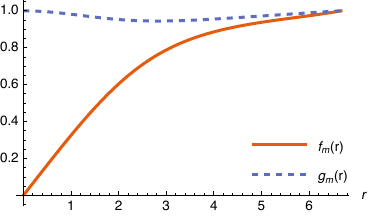}
 \caption{Monopole profile functions $f_m(x)$ and $g_m(x)$.
 }
\label{mprofiles}
\end{figure}

Magnetic monopole solutions in SU(3) have been discussed in Refs.~\cite{Bais:1978yh,Wilkinson:1978zh}.  
For example, the scalar field for the monopole solution
may be written as
\be
\Phi_{\rm m} ({\bf x}) = \eta \left ( f_m (r)  \frac{\sqrt{3}}{2} {\hat r} \cdot {\hat \tau} - g_m(r) \frac{1}{2} T_8 \right )
\label{Phim}
\ee
where ${\hat r}$ is the unit radial vector and ${\hat \tau} = (T_1,T_2,T_3 )$. The monopole
profile functions (for $W_\mu=0$) are obtained numerically and are shown in Fig.~\ref{mprofiles}. Note that $\Phi_m (r=0) \propto -T_8$ while
$\Phi_m (r=\infty) \propto T_8'$ where $T_8'$ is a permutation of the diagonal entries of $T_8$. This implies that $\Tr(\Phi^3)$ has opposite sign within the monopole as compared to the asymptotic value.

The equations of motion are derived from the Lagrangian in \eqref{lagrangian}.
In temporal gauge ($W_0=0$) and, with the inclusion of a damping term these are,
\be
\partial_t^2 \phi^a - D_i D_i \phi^a - \gamma \frac{\phi^b {\dot \phi^b}}{\phi^c \phi^c} \phi^a + \frac{\partial V}{\partial \phi^a} = 0
\label{Phieq}
\ee
\be
[D_\mu , W^{\mu\nu} ] = j^\mu = ig [\Phi, D^\nu\Phi]
\label{Wiaeq}
\ee 
where $\gamma$ is a phenomenological damping parameter that we have introduced 
to account for dissipation such as due to boson decay into fermions. The form of the dissipation term is chosen so that it is consistent with current conservation: $D_\mu j^\mu =0$ (see \cite{Zhang:2019vsb}).

These continuum equations of motion are discretized and the Gauss constraints
are implemented using the technique described in Ref.~\cite{Vachaspati:2016abz}.

Thermal initial conditions for the simulation are implemented by randomly drawing 
the Fourier coefficients for the fields (labeled by ${\vec k}$) from a Gaussian distribution with variance given by the Bose-Einstein (BE) distribution,
\be
\sigma_k^2 = \frac{2}{\omega_k} \frac{1}{e^{\omega_k/T} - 1}
\label{variance}
\ee
The BE 
distribution involves two parameters, one is the temperature, $T$, and the other is
the mass parameter in the dispersion relation $\omega_k = \sqrt{k^2+m^2}$. For the 
purpose of the initial conditions we choose $m=1=T$ in units of the VEV for the scalar fields and set $m=0$, $T=1$ for the gauge fields. We start with 
``half thermal'' conditions since we sample a random distribution of the fields $\Phi$,
$W_i^a$ but not of their time derivatives in order to maintain the Gauss constraints
at the initial time. To compensate for the ``half'' thermal nature, we have included a factor of 2 in the variance in \eqref{variance}. Once we have the fields, we evolve them according to the
equations of motion in \eqref{Phieq}, \eqref{Wiaeq}. The interactions among the
fields result in their thermalization (at a temperature different from $T=1$) after a short time 
which we take to be when the kinetic and gradient energies have stabilized to a constant value. Note 
that the field interactions effectively give a thermal mass to the $\Phi$ field and
there is no need to include a thermal term in the potential during evolution. To then obtain symmetry
breaking, we cool the system by turning on a damping term with damping coefficient
$\gamma =0.6$ in \eqref{Phieq}.

\begin{figure*}
\includegraphics[width=0.31\textwidth,angle=0,trim={1.8cm 1.2cm 3.5cm 1.5cm},clip]{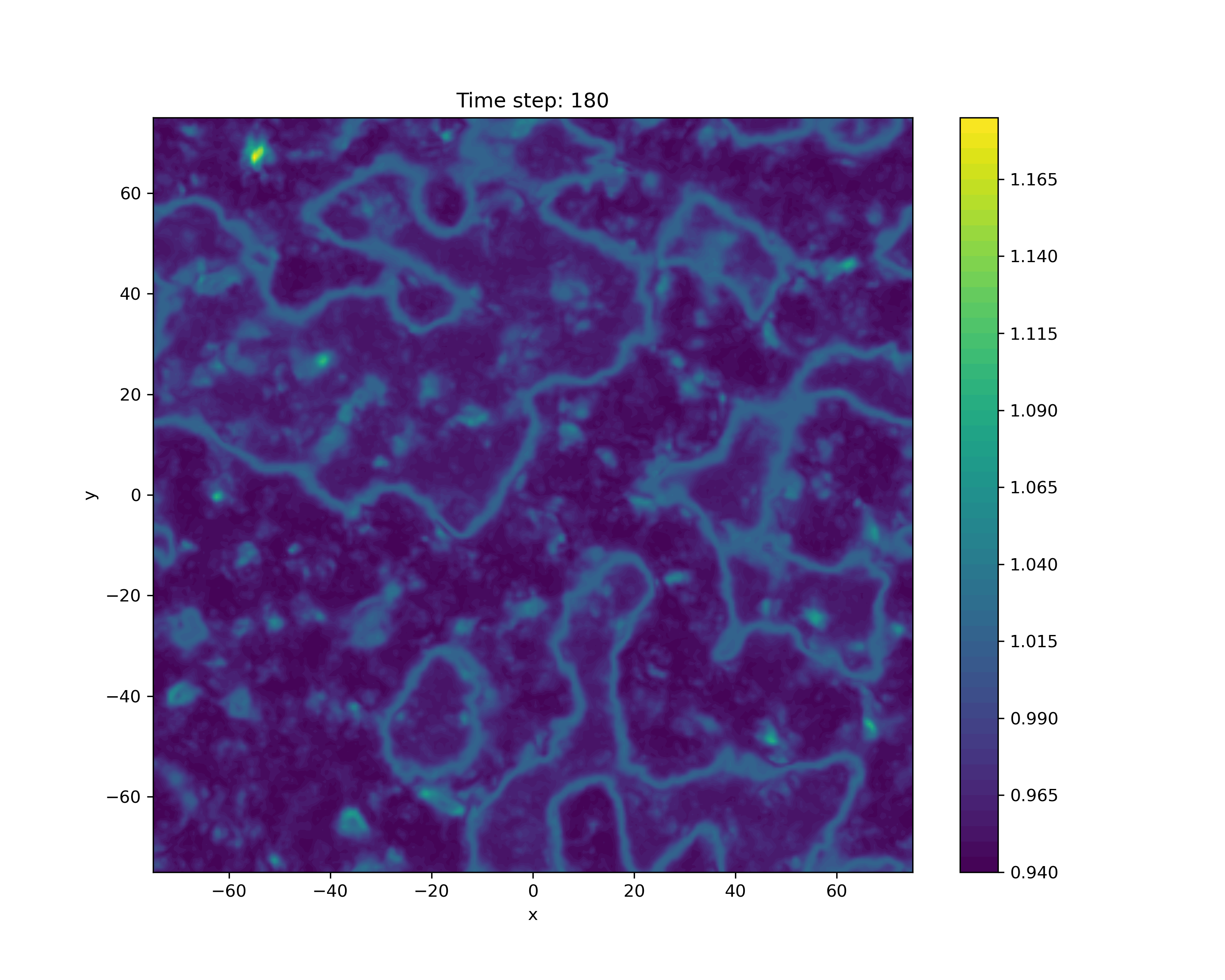} \quad \;
\includegraphics[width=0.31\textwidth,angle=0,trim={1.8cm 1.2cm 3.5cm 1.5cm},clip]{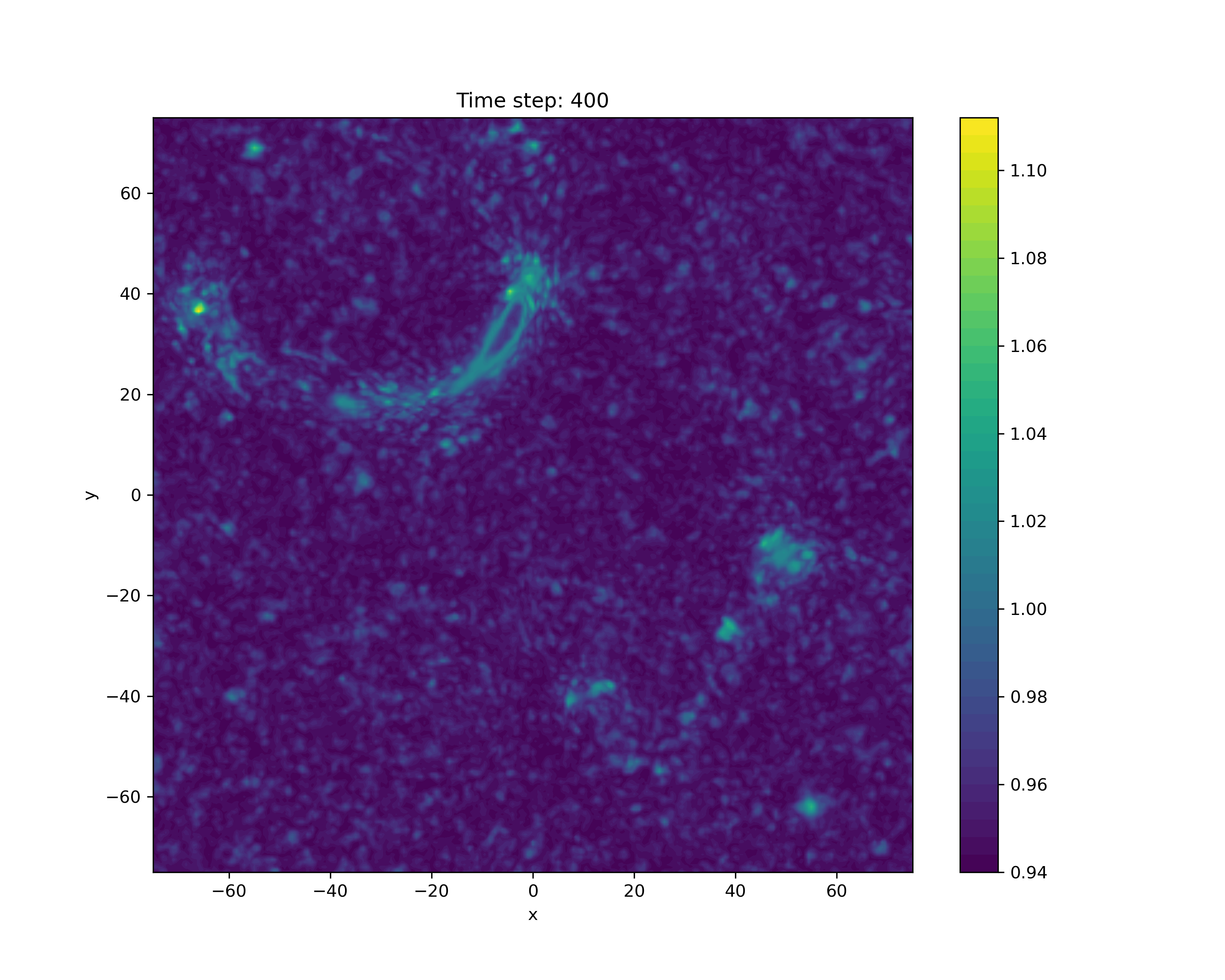} \quad \;
\includegraphics[width=0.31\textwidth,angle=0,trim={1.8cm 1.2cm 3.5cm 1.5cm},clip]{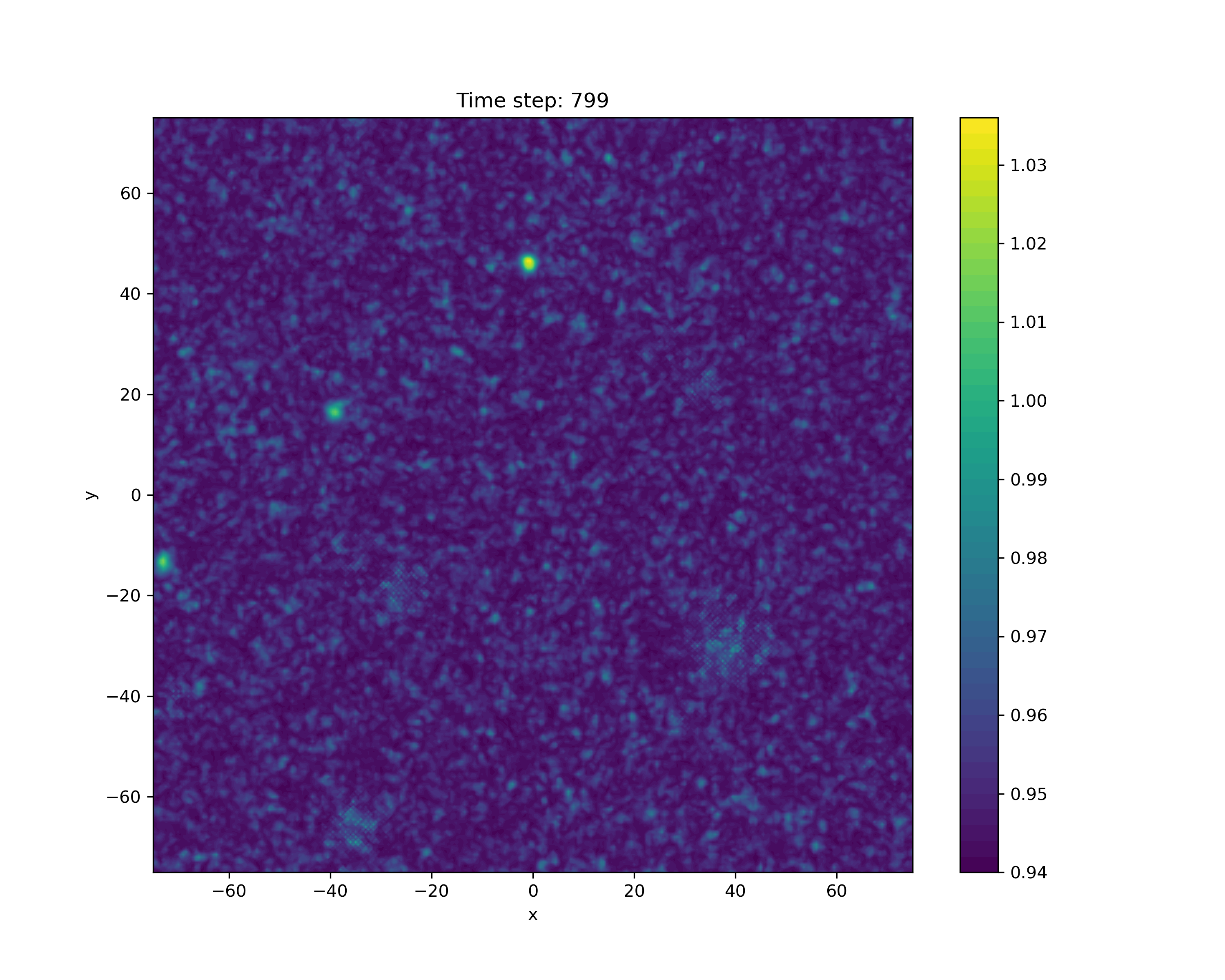}
 \caption{Potential energy density on the $z=0$ slice at three intermediate time steps in the simulation
 with $\epsilon = 0.02$, showing the presence of domain walls and a few surviving monopoles.
 }
\label{PEFig}
\end{figure*}

After symmetry breaking we look for a network of domain walls and a distribution of magnetic monopoles. A simple diagnostic for domain walls is to look for sign changes of $\Tr (\Phi^3)$ across links in the lattice. The issue is that
$\Tr (\Phi^3)$ also changes sign within the magnetic monopole. In other words, the magnetic monopole is like a very small spherical domain wall that carries magnetic charge~\cite{Pogosian:2001fm}. To distinguish between monopoles and walls, we define the monopole to be spatially localized to a small volume away from extended domain walls and to have non-trivial topological winding\footnote{The algorithm in Ref.~\cite{Ng:2008mp} to find SU(3) monopoles assumes that $\Phi$ is strictly in its vacuum manifold. This assumption is valid at very late 
times but not at earlier times when there can be significant fluctuations around the vacuum.}. 

We start looking for monopoles once the average value of $\Tr(\Phi^2)$ over the lattice reaches $0.5$, indicating that most of the field has settled close to its true vacuum where $\Tr(\Phi^2) = 0.64$. 
Then we locate the points where $\Tr(\Phi^2)$ is below $0.3$ as this indicates regions where the potential energy is high and likely correspond to locations of defects. We identify points of local minima of $\Tr(\Phi^2)$ by comparing their value at each point with neighborhood points centered around a ($3^3$) sub-lattice. We construct small ($7^3$) sub-lattices around such points -- large enough to encompass most of the inner core of the monopole -- and check if $\Tr(\Phi^3)$ is the same sign on the corners of the sub-lattice. This tells us that a domain wall is not passing through the sub-lattice. In this case the sign of $\Tr(\Phi^3)$ on the boundary determines the sign of the $Z_2$ vacuum. We then diagonalize $\Phi$ within the sub-lattice and determine the functions $f_m$ and $g_m$ in \eqref{Phim} by solving the cubic equations given by,
\be
\Tr (\Phi_m^2) = \frac{\eta^2}{8}(3 f_m^2+g_m^2 ), \ \ 
\Tr (\Phi_m^3) = \frac{\eta^3}{32\sqrt{3}} g_m (9 f_m^2 - g_m^2) \nn
\ee
A unique solution for $g_m$ and $f_m$ is obtained by using the condition that the sign of $g_m$ is the same as the sign of the $Z_2$ vacuum (sign$(g)=\pm$ for $\Phi \propto \pm T_8$ in the vacuum). Then we find the locations within the sub-lattice where $f_m^2$ is below a threshold ($f^2 < 0.08$). Such points are close to the center of a monopole (see Fig.~\ref{mprofiles}). Denote such a  point by ${\bf x}_*$. Let $U_* \in \SUthree$ diagonalize ${\tilde \Phi} ({\bf x}_*)$, where the tilde denotes $\Phi$ multiplied by the sign of the $Z_2$ vacuum. Denote the diagonal matrix as $D_* \equiv U_* {\tilde \Phi}_* U_*^\dag$.
The diagonal matrix $D_*$ has two eigenvalues that are smaller than the third as it is approximately
proportional to $T_8$ (see \eqref{Phim}) or to $T_8$ but with permuted diagonal entries. These smallest eigenvalues identify the unbroken SU(2) block.
We denote the generators of the SU(2) group that commute with $D_*$ by
${\vec \Sigma}$. Then the generators of the unbroken SU(2) group are 
${\vec \tau} \equiv U_*^\dag {\vec \Sigma} U_*$. Once we have ${\vec \tau}$, we
construct the radial unit vector in \eqref{Phim} in the sub-lattice around ${\bf x}_*$,
\be
{\hat r}_{i,j,k} = \Tr ( {\vec \tau} \Phi_{i,j,k})/ | \Tr ( {\vec \tau} \Phi_{i,j,k})|
\ee
In a given cell, we now have 8 ${\hat r}$ vectors, one at each vertex of the cell.
These 8 ${\hat r}$ vectors define a mapping from the surface of the cell to a two
sphere. The next step is to determine if the mapping is topologically 
non-trivial~\cite{Leese:1990cj}.

Consider a ``triangular plaquette'' of the cell, for example formed by the vertices
$(i,j,k)$, $(i+1,j,k)$ and $(i,j+1,k)$. The ${\hat r}$ vectors at these points maps
the triangular plaquette to a spherical triangle on the two sphere. The area of this 
spherical triangle is given by
\be
{\cal A} = 2\, \tan^{-1} \left ( \frac{{\hat r}_1 \cdot {\hat r}_2 \times {\hat r}_3}{1+r_{12}+r_{23}+r_{31}}
 \right )
\ee
where $r_{ab} = {\hat r}_a \cdot {\hat r}_b$ and $a,b =1,2,3$ refer
to the three vertices of the spherical triangle and the triangular plaquette $abc$ is 
oriented with areal vector pointing out of the cell.
Three points on a two sphere define two spherical triangles and we
always take $A$ to be the area of the smaller triangle: $-2\pi \le A \le 2\pi$. 
Next we sum the areas of the 12 spherical triangles obtained
as maps from the 12 triangular plaquettes (``tp'') of the cell to get the topological
winding
\be
w = \frac{1}{4\pi} \sum_{\rm tp} A_{\rm tp}.
\ee
If $w=+1$ ($w=-1$) there is a monopole (antimonopole) in the cell. 

The algorithm for finding monopoles works better at late times when $\Phi$ is nearly settled in its true vacuum. There can still be some glitches, especially at early times or when monopoles come close together.

Once we have determined all the sub-lattices that contain magnetic monopoles, we turn to domain walls. These are detected simply by changes of $\Tr (\Phi^3)$ in all cells that are not in sub-lattices containing magnetic monopoles.

Our simulations are performed on $300^3$ cubic lattices with periodic boundary conditions,
with lattice spacing $dx=0.5$ and time step $dt = dx/3$.
Snapshots of the potential energy density at three different times on the $z=0$
 slice are shown in Fig.~\ref{PEFig}. We see both an evolving network of domain walls
 and a few monopoles. At early times, monopoles do not necessarily follow the solution in \eqref{Phim} and there are fluctuations in the number counts, but our algorithm does well at later times, as $\Phi$ settles into its vacuum.

 \begin{figure}
\includegraphics[width=0.4\textwidth,angle=0]{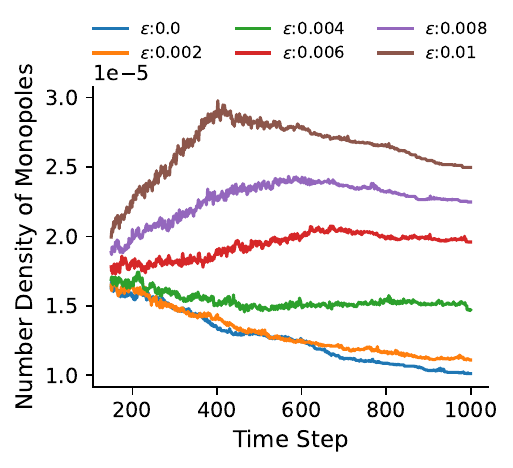}
 \caption{Number density of monopoles as a function of time for several values of $\epsilon$.
 }
\label{Nvst}
\end{figure}

In  Fig.~\ref{Nvst} we plot the time evolution of the number density of monopoles after
averaging over 10 random realizations. In the run with the strongest bias $\epsilon=0.01$, the domain walls annihilate fastest.
The run with $\epsilon=0$ has topological domain walls which are present by the end of the simulation. The plots cannot be trusted at early times ($t \lesssim 200$) since $\Phi$ is still far from its vacuum expectation value.
At late times, after the domain walls have annihilated,
we expect the time evolution to follow previously obtained results
~\cite{Zeldovich:1978wj,Preskill:1979zi,Martins:2008zz,Sousa:2017wvx,Hindmarsh:2025vxh}. (In a cosmological setting, the monopole number density is expected to freeze out.)
However, at the times we are considering, the evolution shows more rapid annihilation
due to domain walls with a suppressed final number density of monopoles.

When a domain wall with magnetic charge collapses it will form a magnetic monopole. This is one way how the monopoles are produced in our simulation. When uncharged domain walls collapse, they can also form monopole-antimonopole pairs very close to each other. These pairs then quickly annihilate, which shows up as fluctuations in Fig.~\ref{Nvst} at the time steps when the domain walls collapse.

Our results clearly show that a decreasing bias, $\epsilon$, leads to a smaller final number
density of monopoles.
This is to be expected in the sweeping scenario since smaller $\epsilon$ implies a greater 
areal density of domain walls and the walls also survive longer. At very small $\epsilon$, the 
density of domain walls at early times in our simulations is very high and this also
 suppresses the formation of monopoles. 
 
The sweeping scenario has been discussed for minimal SU(5) Grand Unification in the cosmological context in Ref.~\cite{Dvali:1997sa}. Constraints arise because the walls have to survive long enough to sweep up the monopoles but not survive so long that the universe becomes domain wall dominated and violates cosmological constraints. In terms of the $Z_2$ violating parameter, $\epsilon$ should be small enough for efficient sweeping but large enough that the domain walls annihilate prior to big bang nucleosynthesis. While there is no direct mapping between the SU(5) parameters in~\cite{Dvali:1997sa} and our SU(3) parameters, based on the final constraint in~\cite{Dvali:1997sa} we expect that the parameter window where sweeping can solve the monopole problem to be
 \be
 (M_{\rm GUT}/M_P)^2 \eta \lesssim \epsilon \lesssim 10 \lambda \eta.
 \ee
 where $M_{\rm GUT}$ is the energy scale of the GUT and $M_P$ is the Planck scale.
 The parameter range in Ref.~\cite{Dvali:1997sa} was obtained by 
 requiring that the domain walls decay prior to dominating the universe. In principle there could be a period of domain wall domination if it is consistent with cosmological observations and this would widen the parameter range. 
 
On the other hand, we have observed several instances in our simulations where a closed domain wall collapses to form a magnetic monopole. This process is quite natural since the domain walls sweep the monopoles, acquire magnetic charge, and then collapse. The net magnetic charge on a closed domain wall may not vanish and when it does collapse, it will lead to the production of magnetic monopoles.
If the closed domain wall is large and carries magnetic charge, it may collapse to a magnetically charged black hole~\cite{Vachaspati:2017hjw,Ferrer:2018uiu} leading to a smoking gun signature of GUTs and the sweeping scenario. 
Roughly we expect a (spherical) domain wall of size $R$ and mass $M \sim \sigma 4\pi R^2$, where $\sigma$ is the wall energy density, to have captured $N = n_m 4\pi R^3/3$ monopoles and antimonopoles, where $n_m$ is their number density. The magnetic charge within the domain wall is given by a surface integral (by Gauss' law) and the net magnetic charge, taking into account fluctuations of the surface integrand, grows $\propto R$~\cite{Vachaspati:1984dz}. 
If the wall collapses into a black hole the magnetic charge on the black hole will be related to its mass by $Q_m \propto e_m \sqrt{M}$ where $e_m$ is the magnetic charge of a monopole. 
If the black holes have mass below $10^{15}\,$g, they will evaporate until they reach extremality, contributing to the dark matter in the form of extreme magnetic Reissner-Nordstrom black holes and with interesting standard model effects in their vicinity~\cite{Maldacena:2020skw}.
We plan to investigate this scenario in more detail in the near future.

 One difference we can expect between the cosmological evolution and our flat space results is that the expansion of the universe causes the domain wall area within a Hubble patch
 to grow like $t^2$ where $t$ is the cosmic time during the period the bias term is unimportant for the wall dynamics. This feature will lead to more efficient sweeping. Unlike in our simulations, cosmological walls never completely straighten out and have relativistic velocities. Once the
 bias term becomes important for the wall dynamics, the collapse of the wall network is expected
 to occur rather quickly even with Hubble expansion.

Another implication of the sweeping scenario is that there should be a stochastic gravitational wave background due to gravitational wave emission from biased domain walls. 
 Since these walls form at the same high energy scale as magnetic monopoles, the gravitational wave background could be significant and may be detectable by pulsar timing arrays~\cite{Hiramatsu:2010yz,Kawasaki:2011vv,Hiramatsu:2013qaa,Saikawa:2017hiv,Ferreira:2022zzo,Kitajima:2023cek,Blasi:2025tmn}.

To summarize, for the first time we have simulated symmetry breaking in GUT inspired models. We find that the outcome for the formation of magnetic monopoles is highly sensitive to the presence of biased domain walls. The domain walls sequester magnetic monopoles with potentially important implications for cosmology.

More plots and animations of our data can be found on \url{https://youtu.be/1_Q4T_3oeDc}.

This work was supported by the U.S. Department of Energy, Office of High Energy 
Physics, under Award No.~DE-SC0019470. Computational resources were provided by the Sol supercomputer 
at Arizona State University \cite{jennewein2023sol}.

\appendix

\bibstyle{aps}
\bibliography{paper}

\end{document}